\documentclass[aps,prl,twocolumn,groupedaddress,floatfix]{revtex4}
  \usepackage{graphicx}  \usepackage{psfrag}
\usepackage{amsfonts,amssymb,amsmath}

\begin{document}
\title{Cycloaddition Functionalizations to Preserve or Control \\ 
the Conductance of Carbon Nanotubes } 

\author{Young-Su Lee and Nicola Marzari} 

\affiliation{Department of  Materials Science and Engineering,
and Institute for Soldier Nanotechnologies, 
Massachusetts Institute of Technology, 
Cambridge, Massachusetts 02139, USA }
                                                                                                                               
\date{\today}
                                                                                                                               
\begin{abstract}
We identify a class of covalent functionalizations that preserves or
controls the conductance of single-walled metallic carbon nanotubes.  [2+1]
cycloadditions can induce bond cleaving between adjacent sidewall
carbons, recovering in the process the $sp^2$ hybridization and the ideal
conductance of the pristine tubes. This is radically at variance with
the damage permanently induced by other common ligands, where a single
covalent bond is formed with a sidewall carbon.  Chirality, curvature,
and chemistry determine bond cleaving, and in turn the electrical
transport properties of a functionalized tube. A well-defined range of
diameters can be found for which certain addends exhibit a bistable
state, where the opening or closing of the sidewall bond, accompanied
by a switch in the conductance, could be directed with chemical, optical 
or thermal means. 
\end{abstract}

\pacs{} \maketitle
\begin{figure}[b]
\centering
\includegraphics[height=0.4\textwidth,angle=270]{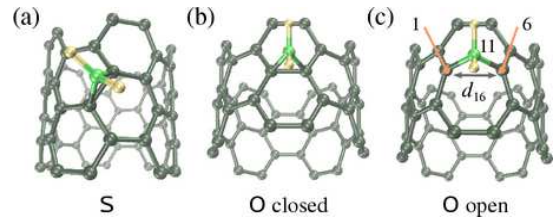}
\caption{  
The three different configurations for a functional group on an armchair nanotube are shown (CH$_{2}$ on
a (5,5) CNT): (a) skewed \textsf{S},
(b) orthogonal \textsf{O} with an intact (``closed'') sidewall bond,  and (c)
orthogonal \textsf{O} with a broken (``open'') sidewall bond.
}
\label{fig:FIG1}
\end{figure}

Chemical functionalizations of carbon  nanotubes (CNTs) are the subject of
intensive  research \cite{Dyke04,LuX05},
and  could offer  new and  promising  avenues to
process and assemble tubes, add sensing capabilities, 
or tune their  electronic properties 
(e.g., doping levels, Schottky barriers, work functions, and electron-phonon couplings).   
However, the benefits of functionalizations are compromised  by 
the damage to the conduction channels that follows
$sp^3$ rehybridization  of  the  sidewall carbons \cite{Kamaras03,LeeY05},  
as evidenced by absorption spectra and electrical   transport
measurements \cite{Strano03,WangC05,Klinke06}.
We report here on a  class of cycloaddition functionalizations that 
preserves instead the remarkable transport properties of metallic CNTs.
In addition, we identify a subclass of  addends that displays a
reversible valence tautomerism that can directly control the conductance.
\\ 
We focus here on [2+1] cycloaddition reactions, where
the addition of a  carbene or a
nitrene  group  saturates  a  double-bond between  two  carbon  atoms,
forming  a  cyclopropane-like  three-membered  ring.   Such
functionalizations   have    been   reported   extensively    in   the
literature \cite{ChenJ98,Holzinger01,Coleman03}.   
All our calculations are performed
using density-functional theory
in the Perdew-Burke-Ernzerhof 
generalized gradient approximation (PBE-GGA) \cite{Perdew96},
ultrasoft pseudopotentials \cite{Vanderbilt90},
a planewave basis set with a cutoff of 30 Ry for the wavefunctions
and 240 Ry for the charge density,
as implemented in \textsc{q}{\small uantum}-\textsc{espresso} \cite{ESPRESSO}.
\\
We examine first
the simplest members in this class of addends, CH$_2$ and NH.
Fig. \ref{fig:FIG1}(a)
and  (b)  show the  two  inequivalent  choices  available on  a  (5,5)
metallic  CNT; for  convenience, we  label these
as   ``\textsf{S}'' (skewed) and
``\textsf{O}'' (orthogonal),
reminiscent of the relative positions of the sidewall carbons with respect
to the tube axis.
Our simulation cells
include 12 $n$ carbons for a given ($n$,$n$) CNT, plus one functional group.
We use a 1$\times$1$\times$4 Monkhorst-Pack mesh (including $\Gamma$)  
for structural optimizations,
and a 1$\times$1$\times$8 mesh for single-point energy calculations,
with a cold smearing of 0.03 Ry \cite{Marzari99}.
\begin{figure}[t]
\centering \includegraphics[width=0.4\textwidth]{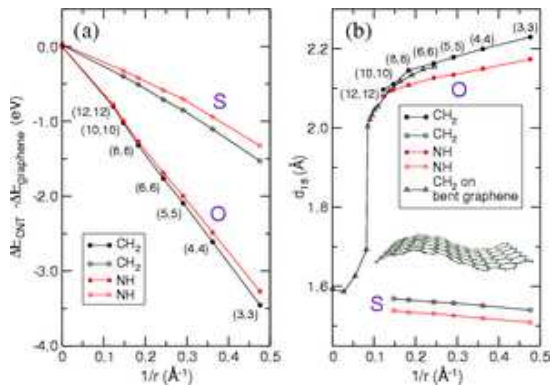}
\caption{
(a) Energy change $\Delta E_{CNT}$ upon functionalization as a function of
curvature, for ($n$,$n$) armchair CNTs - the zero reference is for graphene.
(b)  Sidewall equilibrium bond distance C$_1$-C$_6$ ($d_{16}$) 
for ($n$,$n$) CNTs and for a
bent graphene sheet, functionalized  at the stable \textsf{O}  site.  The sidewall bond is
broken  in all the ($n$,$n$) CNTs considered. Continuous bending of a graphene sheet shows that a
well-defined transition  from the closed to  the open form takes place as
the  curvature increases.
}
\label{fig:FIG2}
\end{figure}
\\
First,  
and for  ($n$,$n$) metallic tubes,
we  highlight how  strongly  the  reaction  energies of  these
functionalizations depend on the curvature of the nanotubes, and on their
attachment  sites \textsf{S} and  \textsf{O}.
We plot in Fig. \ref{fig:FIG2}(a)
the  reaction  energies $\Delta  E_{\text{CNT}}$  (defined as  $\Delta
E_{\text{CNT}} = E_{\text{CNT-func}} - E_{\text{CNT}}-E_{\text{func}}$, func=CH$_2$ or NH),
taking as a
zero reference  the same reaction on graphene.   The reaction energies
have   a  well-defined   linear  dependence   on   curvature,  clearly
demonstrating  the higher reactivity  of smaller-diameter  tubes.  The
\textsf{O} site  is always more  stable, and significantly so  for all
diameters considered; at room temperature all small-diameter armchair CNTs
strongly favor the \textsf{O} configuration.
The energy difference between the \textsf{O} and the \textsf{S} form of the (5,5) CNT
is 1.24 eV, which is 
in good agreement with other
planewave basis calculations (1.24 eV, Ref. \cite{LuJ05})
and localized basis calculations (1.4 eV, Ref. \cite{Bettinger06}). 
\\
Second, we find that the  C$_1$-C$_6$ distance for  this \textsf{O}
configuration ($d_{16}$ in  Fig. \ref{fig:FIG1}(c)) is much larger than the
usual C-C distance (1.54 $\text{\AA}$ in diamond and 1.42 $\text{\AA}$ in graphite),
a clear indication  that the sidewall bond  is broken
\cite{ChenZ04,ZhaoJ05,LuJ05,Bettinger06}.
For  the   CH$_2$   and  NH
cycloadditions,
we observe bond cleaving for all nanotubes studied
(up to (12,12)); on the other hand, 
in graphene the bond is intact.  
We estimate the critical diameter
that separates the two regimes
by bending a graphene sheet:
we    can see   in
Fig. \ref{fig:FIG2}(b) that  such model 
closely reproduces the nanotube results,
and a sharp transition from
the bond-intact to the bond-broken form
takes place  around  a diameter  of  2.4 nm, 
i.e. a (18,18) tube, for  CH$_2$  functionalizations.
\\
Broken or intact sidewall bonds play a fundamental role in the
electronic transport  properties  of a metallic nanotube. 
We show this in Fig. \ref{fig:FIG3}, where the Landauer conductance 
is calculated for a 
(5,5) CNT functionalized 
with dichlorocarbene (CCl$_2$) (first reported 
experimentally in 1998 \cite{ChenJ98}),
using an approach 
recently introduced by us that allows to treat 
realistic nanostructures with thousands of atoms while
preserving
full  first-principles  accuracy \cite{LeeY05}.
For this diameter the sidewall bond is broken, and the scattering 
by a single CCl$_2$ group is found to be remarkably weak, 
with the  conductance approaching its  ideal value.
Even after  adding 30
groups  on the  central 43 nm segment of  an infinite  nanotube
the conductance is only reduced by 25\%.
This is in sharp contrast
with the case of a hydrogenated tube,
where the conductance
drastically drops practically to zero when functionalized with
a comparable number of ligands.
This result is easily rationalized. 
Hydrogen and other single-bond covalent ligands induce
$sp^3$ hybridization of the sidewall carbons, and these chemical defects
act as very strong scatterers \cite{LeeY05}.
Such dramatic decrease in the conductance has also been recently confirmed 
experimentally \cite{Klinke06}.
On the other hand, after bond cleavage C$_1$ and C$_6$ 
recover a graphite-like bonding environment (Fig. \ref{fig:FIG1}(c)),
with three covalent bonds to their nearest neighbors. Their electronic
orbitals go back to $sp^2$ hybridization, allowing for 
the $p_z$ orbitals of C$_1$ and C$_6$ 
to be recovered (as confirmed by inspection of the
maximally-localized Wannier functions \cite{Marzari97})
and to contribute again to the graphitic
$\pi$ manifold. 
The net result is that conductance approaches again 
that of a pristine CNT, highlighting the promise
of cycloadditions in {\it preserving} the conductance 
of metallic nanotubes.
\\
In principle, around the 
critical diameter shown in Fig. \ref{fig:FIG2}(b), a 
functional group could  be found that displays as stable states 
on the same tube both the open and the closed configuration.
If the case, interconversion 
between the two valence tautomers would have a direct effect
on the conductance. We illustrate this paradigm with the dichlorocarbene example
of Fig. \ref{fig:FIG3}; if we force the tube in its closed configuration,
with all the sidewall bonds frozen in their ideal pristine-tube geometry,
conductance decreases by a factor of 2 or 3 with respect to the case of the relaxed
tube \cite{comment1}. 
The configuration where the sidewall bonds are intact is not stable
for this case, but, depending on the chemistry of the addends and the diameter considered,
optimal ligands could be found for which a double-well stability  is present.
\begin{figure}[t]
\centering
\includegraphics[height=0.40\textwidth,angle=270]{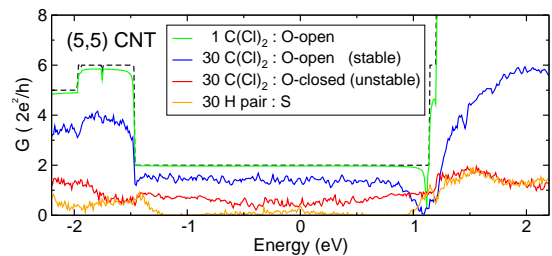}
\caption{ Quantum  conductance for
a (5,5) CNT functionalized with one CCl$_2$ group,
with 30  CCl$_2$ groups, and with 30 hydrogen pairs (dashed line: pristine (5,5) CNT).
A CCl$_2$ addend will choose the \textsf{O}-open configuration, while hydrogens prefer to
pair in the \textsf{S} configuration;
the results corresponds to these stable choices.
The conductance for the energetically-unstable \textsf{O}-closed configuration
is also plotted. 
The 30 functional groups are arranged randomly on the central 43-nm segment
of an otherwise infinite tube; the conductance is then averaged over 10 different configurations.}
\label{fig:FIG3}
\end{figure}
\\
In order to identify
the factors that determine stability in the open or closed form,
we first screen several addends on small molecules.
We find that the  bridged 1,6-X-[10]annulene (inset of Fig. \ref{fig:FIG4}(a))
is an excellent molecular homologue  of a functionalized CNT \cite{ChenZ04}.
\begin{figure}[t]
\includegraphics[width=0.48\textwidth]{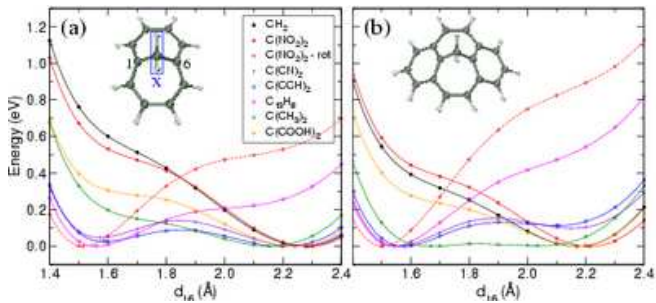}
\caption{  
Potential energy  surface as a  function  of $d_{16}$  for
select cases of (a) \textbf{1} and
(b) \textbf{2}.
C(CN)$_2$ (violet) and C(CCH)$_2$ (blue)
show a double-well minimum in both \textbf{1} and \textbf{2}.
C(NO$_2$)$_2$-rot (dashed red) indicates
the unstable conformation where the two NO$_2$ groups are rotated 
from their equilibrium position (solid red) by 90$^\circ$.
}
\label{fig:FIG4}
\end{figure}
It is well-known that the substitutional group  X dictates
the preference  for  the annulene (henceforth labelled as \textbf{1o})
or for  its  valence tautomer, a bisnorcaradiene derivative (\textbf{1c}),
corresponding to the open and  closed configurations of a functionalized CNT
\cite{Bianchi73,Guenther75,Pilati76,Bianchi80,Bianchi81,Vogel82}.
Similar  tautomerization 
between an open (\textbf{2o}) and a closed (\textbf{2c}) form 
takes place in a pyrene derivative \cite{Wirz84}
(inset of Fig. \ref{fig:FIG4}(b)).
We thus tested on these molecules the substitutional groups X= CH$_2$,   NH,   SiH$_2$,   C(NO$_2$)$_2$,
C(CN)$_2$,   C(CCH)$_2$,  C(CH$_3$)$_2$,  C(COOH)$_2$,
CCl$_2$, C(NH$_2$)$_2$,
C$_6$H$_4$O \cite{Eiermann95}, and C$_{13}$H$_8$ \cite{Eiermann95}.
To assess the accuracy of our PBE-GGA approach,
we compare in Table \ref{table:Table1}
our results for $d_{16}$ in \textbf{1} with 
those obtained from experiments or
other theoretical methods (second order M\o ller-Plesset perturbation theory (MP2) \cite{Moller34}
and Becke three parameter Lee-Yang-Parr hybrid functional (B3LYP) \cite{Becke93}), finding
excellent agreement.
Hydrogen and halogens have been reported to
stabilize  \textbf{1o} both  experimentally and  theoretically
\cite{Bianchi80,Pilati76,ChoiC98}. 
\squeezetable
\begin{table}[b]
\caption{
Experimental and theoretical
$d_{16}$ of {\textbf 1}.
Note that the calculations assume isolated molecules at 0 K,
while experimental data are obtained from crystalline systems 
at finite temperature.
Theory predicts two stable minima for X=C(CN)$_2$. 
The long $d_{16}$ of X=C(CH$_3$)$_2$ indicates that
the potential energy surface would be very flat,
which is also predicted in Fig. \ref{fig:FIG4}, especially for {\textbf 2}. 
}
\label{table:Table1}
\begin{center}
\begin{tabular}{l|cccc}
\hline
X&Expt.&MP2&B3LYP&This work\\
\hline
CH$_2$        & 2.235\footnotemark[1]   & 2.251\footnotemark[6]            & 2.279\footnotemark[7]            & 2.278 \\
CF$_2$        & 2.269\footnotemark[2]   & 2.268\footnotemark[6]            & 2.296\footnotemark[6]            & 2.300 \\
C(CN)$_2$     & 1.542\footnotemark[3]   & 1.599, 2.237\footnotemark[6] & 1.558, 2.253\footnotemark[8] & 1.572, 2.245 \\
C(CH$_3$)$_2$ & 1.836\footnotemark[4]   & 2.156\footnotemark[6]            & 2.168\footnotemark[6]            & 2.151 \\
NH            & (open)\footnotemark[5]  &                      & 2.237\footnotemark[7]            & 2.239 \\
\hline
\end{tabular}
\end{center}
$^a$Ref. \cite{Bianchi80},
$^b$Ref. \cite{Pilati76},
$^c$Ref. \cite{Vogel82},
$^d$Ref. \cite{Bianchi73},\\
$^e$Ref. \cite{Vogel65},
$^f$Ref. \cite{ChoiC98},
$^g$Ref. \cite{JiaoH02},
$^h$Ref. \cite{Gellini00}
\end{table}
Cyano  group favors  \textbf{1c} experimentally
(the possibility of coexistence with \textbf{1o} is discussed) \cite{Vogel82},
and two minima have been actually predicted theoretically \cite{ChoiC98,Gellini00}.
\\
We show the  potential energy profile
of select groups in Fig. \ref{fig:FIG4}. 
All carbenes stabilizing the closed form in both \textbf{1} and \textbf{2}
share a common feature: 
partially-occupied $p$ orbitals parallel to
the in-plane 
$p_{\sigma}$ orbital of the bridgehead C$_{11}$ atom \cite{Mealli97}
(the plane considered is that of C$_1$-C$_{11}$-C$_6$ of Fig. \ref{fig:FIG1}(c)).
This conclusion is strongly supported by examining the
energy minimum conformation for X=C(NO$_2$)$_2$.
At equilibrium, the two oxygen atoms in the NO$_2$ group
lie on a line parallel to the C$_1$-C$_6$ bond,
and the open form is stable (solid red in Fig. \ref{fig:FIG4}).
The sidewall bond will
switch from open to closed upon rotation of
the two  NO$_2$  groups   by 90$^{\circ}$  
thereby placing the 
$p$ orbitals of the nitrogens parallel to the
$p_{\sigma}$ orbital of C$_{11}$ (dashed red in Fig. \ref{fig:FIG4}). 
Among   all  the
substituents screened, we find that SiH$_2$,  C(CCH)$_2$ and  C(CN)$_2$ show
most clearly the  presence of two minima in their potential energy surface;
we choose here X=C(CN)$_2$ as the most promising candidate
since
both \textbf{1c} \cite{Vogel82} and 
a C$_{60}$ derivative \cite{KeshavarzK96} have already been synthesized. 
\\
We explored, therefore, the potential energy landscape for armchair CNTs
in the case of C(CN)$_2$ cycloadditions.
The  results shown 
in Fig. \ref{fig:FIG5} 
reflect closely those found in the molecular homologues.
\begin{figure}[t]
\centering \includegraphics[height=0.25\textwidth,angle=270]{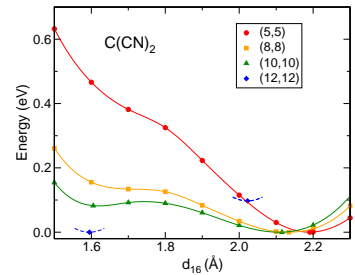}
\caption{ 
Potential energy surface for ($n$,$n$) CNTs functionalized
with C(CN)$_2$. Both (10,10) and (12,12) CNTs display a double-well minimum.
}
\label{fig:FIG5}
\end{figure}
A unique minimum in the open form is found in
small-diameter tubes, as is generally the case in these [2+1]
cycloadditions. As the diameter is increased,
the signature of the closed minimum starts
to appear, first as
an inflection  point ((5,5)  CNT), then as  a local  minimum for the (10,10) CNT 
($\phi$=1.36 nm, as in \textbf{1c}),  and finally as  a
global minimum for the (12,12) CNT ($\phi$=1.63 nm, as in \textbf{2c}).
As discussed before, the conductance 
is controlled by the bonding and hybridization of the sidewall carbons.
We compare in Fig. \ref{fig:FIG6} the  two stable open and closed  
forms for the (10,10) CNT functionalized with C(CN)$_2$.
The  scattering induced by a single group is negligible,
especially in the open form, and
the conductance around  the Fermi energy  is extremely close to
its  ideal value  (Fig.  \ref{fig:FIG6}(a)).
As  the number  of
functional  groups  is increased, the  difference  between  the two 
minima becomes rapidly apparent (Fig. \ref{fig:FIG6}(b)).
\\
Two conclusions can be drawn:
First, even with a large number of
functional groups, the conductance of the tube 
is well preserved, whenever cycloaddition breaks the sidewall bond.
Second, a subclass of substituents can be found 
(e.g., C(CN)$_2$) that stabilize two tautomeric forms on the same tube,
separately displaying high and low conductance.
Several mechanisms, including photochemical, electrochemical, and thermal,
could then direct interconversion
between the two tautomeric forms.
Photochemical  and electrochemical  interconversion rely  on  the fact
that the energy levels of the frontier orbitals are 
affected by $d_{16}$, depending on their symmetries and charge distributions
(e.g., the bond weakens as a filled orbital that has bonding character
along C$_1$-C$_6$ is emptied) \cite{Mealli97,ChoiC98}.
Both  photochemical   excitations  or  electrochemical   reduction  or
oxidation  can populate  or depopulate  those frontier  orbitals that
favor the open or closed form; as a result, they would
modulate the bond distance, and ultimately the conductance. As a proof
of principle, time-dependent density functional calculations in \textbf{2} for X=C(CN)$_2$
show that the first singlet excitation (S$_1$) drives the
system from the open to the closed form \cite{LeeY07}.
A similar conclusion is drawn from experimental observations in
\textbf{2o} for X=CH$_2$, where this open form is
stable in the ground state while the closed
\textbf{2c} is
presumed to be stable in the  S$_1$ energy surface \cite{Wirz84}.
Temperature also plays an  important role, and 
$^{13}$C  NMR spectroscopy and X-ray  data have captured
the  temperature-dependent equilibrium  of similar  fluxional systems:
a higher temperature  stabilizes \textbf{1o}  for X=C(CN)(CH$_3$),
while destabilizing  it  for X=C(CH$_3$)$_2$
\cite{Guenther75,Bianchi81}.
\begin{figure}[t]
\centering 
\includegraphics[height=0.40\textwidth,angle=270]{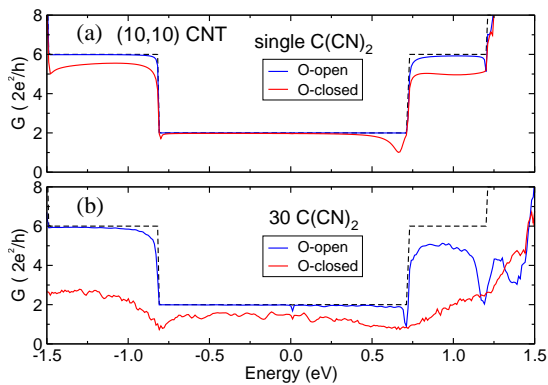}
\caption{  
Quantum   conductance
for a (10,10) CNT functionalized with C(CN)$_2$, in the \textsf{O}-open and \textsf{O}-closed
stable configurations (dashed  line: pristine (10,10) CNT).
(a) Single group. (b) 30 functional groups arranged randomly on the central 32-nm segment
of an infinite tube; the conductance is then averaged over 10 different configurations.
}
\label{fig:FIG6}
\end{figure}
\\
In conclusion, our calculations predict that 
1) a  broad class  of cycloaddition  functionalizations on
narrow-diameter nanotubes recovers, as a consequence of bond cleaving, 
the conductance of the original pristine tubes,  allowing for organic
chemistry approaches  to manipulation  and assembly that  preserve the
remarkable  electronic properties of  these materials,  and 2)  that a
subclass  of  addends, exemplified  in  this  work by  dicyanocarbene,
exhibits fluxional behavior  that could  be controlled  with  optical or
electrochemical   means.  
Such conductance control, if realized, could have practical applications
in nano- and opto-electronics, chemical sensing, and imaging.
\\
The authors would like  to thank Francesco Stellacci (MIT)
and Maurizio Prato (University of Trieste)
for helpful discussions.
This  research has been
supported by MIT Institute  for Soldier Nanotechnologies (ISN-ARO DAAD
19-02-D-0002) and the   National  Science   Foundation
(DMR-0304019); computational facilities have been provided through
NSF (DMR-0414849) and PNNL (EMSL-UP-9597).


\begin{thebibliography}{35}
\bibitem{comment1}
{\bibinfo{journal}{
A recent study that
considered a (6,6) CNT functionalized with CCl$_2$ did not
find that the open configuration displays higher conductance \cite{ParkH06};
we ascribe this discrepancy to the lack of chemical accuracy in
the model H\"uckel Hamiltonian used}}.
\end{thebibliography}
\end{document}